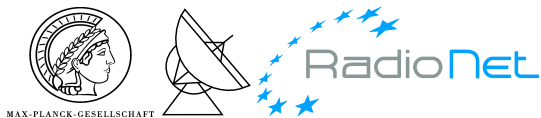

# M 87 and the Dynamics and Microphysics inside the Blazar Zone

P. Hardee[1]

Department of Physics & Astronomy, Gallalee Hall, The University of Alabama, Tuscaloosa, AL 35487, USA

**Abstract.** The blazars provide a considerable opportunity to peer into the workings within a few tens of parsecs of the central engine in AGN. This considerable opportunity involves significant challenges as different macroscopic dynamical processes and microscopic physical processes operating at different locations can be responsible for the observed emission. In this proceedings article I review recent theoretical and numerical results relevant to dynamics inside the blazar zone, review the particle acceleration processes capable of producing the high energy particles required by the observed emission, discuss some of the progress made at the microphysical level, and consider what recent TeV and radio observations of M 87 can tell us about the blazar zone.

## 1. Introduction

The deepest parts of AGN jets are explored by studying the properties of blazars. The spectrum from the radio to the TeV gamma-rays is characterized by two peaks and typically modeled as synchrotron plus inverse Compton emission, e.g., Ghisellini et al. (1998); Sikora (2001). Rapid non-thermal particle acceleration, often assumed to be associated with ejection event driven shocks, is required to generate the synchrotron emission.

Tremendous progress has been made during the past decade on reproducing the spectral energy distributions (SEDs) of blazars. The Fermi/LAT has filled in an important frequency gap between coverage provided by Swift and that provided by ground based Cherenkov arrays. Radio and optical flux, polarization, and timing information along with proper motions, have made it possible to construct likely scenarios for the dynamics and location of emission in the various different wavebands. Nevertheless, much remains to be understood.

The fundamental question is: How is the emission from blazars shaped by (a) the central engine properties, (b) the jet properties, (c) the external environment properties, and (d) in what mixture. Here an ultimate understanding requires that we know: (1) How relativistic jets are launched from the vicinities of super massive black holes (SMBHs). (2) How do the jet structure and composition change along the outflow. (3) What are the particle acceleration processes that produce the TeV particle energies required by the observed emission.

Marscher and coworkers multi-wavelength campaigns provide revealing probes into the structure of the blazar zone (Marcher et al. 2008; Marscher et al. 2010). Here VLBI proper motions supply an estimate of Lorentz factors that link observed timescales to intrinsic timescales and distances between the central engine and the mm radio core. The observed rotation of the optical polarization vector indicates helical twisting of the flow and magnetic field between the central engine and the radio core.

The multi-wavelength observations suggest one or more ejection events resulting in emission from moving jet shocks. The emission comes from internal synchrotron and synchrotron self-Compton (SSC) processes and inverse Compton (IC) processes involving photons from outside the moving zone of emission, e.g., Sikora et al. (1994); Dermer et al. (2009). The SEDs and timing signatures of internal shocks have been considered in some detail by Böttcher & Dermer (2010) for flat-spectrum radio quasars (FSRQs) or low-frequency peaked BL Lac objects (LBLs) where gamma-ray production is dominated by Comptonization of external radiation. For some sources these models may satisfactorily explain timing lags.

Observations suggesting ejection events challenge the assumption of a fixed location for the blazar zone. Still strong outbursts, e.g., 3C 454.3 (Sikora et al. 2008) suggest a blazar zone coincident with a recollimation shock as this fixed structure is more efficient at dissipating energy than internal shocks. Sikora et al. (2008) suggest that a recollimation shock could be located at 3 - 9 pc from the central engine and nearly cospatial with the mm wavelength radio core, e.g., as in Marscher et al. (2008, 2010). Nevertheless, the emission from blazars is likely produced over a broad range of distances from the central engine.

Fitting the SEDs of sources with one-zone homogeneous SSC models makes it possible to find a unique set of parameters for the emitting region (Tavecchio et al. 1998). Application of these models to TeV sources typically requires a Doppler factor, $\delta \equiv [\Gamma(1 - v/c\cos\theta)]^{-1}$, that ranges from 10 - 20 (Katarzynski et al. 2003) up to 50 (Konopelko et al. 2003) depending on the absorption of TeV photons by the IR background. However, these large Doppler and associated Lorentz factors are problematic (Henri & Saugé 2006) because VLBI studies indicate slow moving pc-scale emission features in strong TeV BL Lacs (Lister 2006; Piner, Pant & Edwards 2008).

The slow pc-scale motions led Georganopoulos & Kazanas (2003) to propose that fast moving material at the jet base relativistically boosts radiation produced by slow moving material at the end of a deceleration zone.

Alternatively Ghisellini et al. (2005) proposed a fast jet spine slow jet sheath configuration. In both scenarios the synchrotron and inverse Compton emission come from different regions and this allows reduced Doppler factor values compared to pure one-zone SSC models.

The electron-synchrotron origin of the low frequency emission is well established but the high energy emission could be dominated by cascades if protons are accelerated to p$\gamma$ pion production energies. Such hadronic blazar models in which high magnetic fields are needed for proton acceleration and synchrotron radiation from protons and associated secondaries, e.g., Krawczynski (2007), have not received as much attention as leptonic blazar models. Still there is evidence that leptonically dominated blazar jet models are energetically dominated by protons with Poynting flux to kinetic flux conversion taking place very close to the central engine (Sikora et al. 2005).

Rapid observed variability timescales, $\Delta t$, pose severe constraints on the particle acceleration timescale, $\sim \delta \Delta t$ and the emission region size, $\sim \delta c \Delta t$. Correlated X-ray/TeV gamma-ray flares with timescales from 15 minutes (Mrk 421) to a few hours (Mrk 501 & 1ES 1959+650) have been observed. Interestingly a similar correlation of the radio-to-optical emission with the X-ray or TeV gamma-ray emission is not found, e.g., Krawczynski (2007) and references therein. Thus flares may come from small, a few Schwarzschild radii in size, fast moving "needles" within a slower jet or "jet-in-a-jet" (Levinson 2007; Begelman, Fabian & Rees 2008; Ghisellini & Tavecchio 2008; Giannios, Uzdensky & Begelman 2009).

The X-ray/TeV flares also require very rapid particle acceleration. Fermi acceleration processes are claimed capable of providing the needed acceleration (Henri et al. 1999; Tammi & Duffy 2009), but require large velocity differences combined with turbulent scattering, e.g., shocks or velocity shear along flow boundaries (Stawarz & Ostrowski 2002). Accretion disk shear and turbulence followed by centrifugal acceleration out to the light cylinder is also capable of accelerating protons to $10^{20}$ eV and electrons to the $10 - 100$ TeV required (Istomin & Sol 2009). Fermi processes can take place in super-Alfvénic flow regions but are not likely in the sub-Alfvénic magnetically dominated jet production and acceleration region from which TeV flares might originate. In the absence of Fermi acceleration, acceleration may be provided by vacuum gap electric fields in the black hole magnetosphere (Blandford & Znajek 1977; Krawcznski 2007) or by current driven instability and magnetic reconnection leading to conversion of Poynting flux to kinetic flux (Sikora et al. 2005).

Of course, escape of TeV photons depends on the ambient photon field (Ghisellini & Tavecchio 2009). Here TeV photons might escape from the center of BL Lacs as a result of a reduced ambient photon field that might accompany the reduced accretion rate compared to the FSRQs (Ghisellini et al. 2009) .

## 2. MHD dynamics inside the blazar zone

The location of blazar emission in different wavebands could stretch from the black hole ergosphere to over ten parsecs along the relativistic jets produced by the central engine. Thus, how relativistic jets are launched from the vicinities of SMBHs, and how the jet structure and composition change along the outflow are critical issues for blazars. Theoretically we need to study RMHD processes beginning at scales smaller than the gravitational radius of the accreting black hole, where magnetic and electric fields likely dominate the jet dynamics, out to distances where kinetic effects likely dominate the jet dynamics.

GRMHD codes are used to study the extraction of rotational energy from a spinning black hole, i.e., Blandford-Znajek mechanism (Komissarov 2005; McKinney 2005), and from the accretion disk Meier (2005), i.e., Blandford-Payne mechanism (Blandford & Payne 1982). Jet generation simulations (Hawley & Krolik 2006; McKinney 2006) show development of angular momentum transfer in the accretion disk, leading to diffusion of matter and magnetic field inwards, and unsteady outflows near a centrifugally supported "funnel" wall (Hawley & Krolik 2006). In general, GRMHD simulations with spinning black holes indicate jet production with a Poynting-flux high Lorentz factor spine with $v \sim c$ and $\Gamma \sim 10$, and a matter dominated sheath with $v \sim c/2$ possibly embedded in a lower speed, $v << c$, disk/coronal wind. This basic result suggests the fast jet spine and slower jet sheath type structure advocated to reduce the need for very high Lorentz and Doppler boost factors. We note however that the transverse profile of the Lorentz factor can achieve a maximum in a tenuous boundary layer between a Poynting-flux spine and a dense low speed kinetically dominated sheath (Mizuno et al. 2008; Aloy & Mimica 2008).

Until recently simulation work considered only axisymmetric systems. As a result the efficiency of this mechanism has been questioned because the magnetic configurations are susceptible to a disruptive current driven (CD) kink instability (Begelman 1998). Even so, the outflow may be well collimated by the large-scale poloidal magnetic field of the accretion disk (Spruit, Foglizzo & Stehle 1997). In this scenario the CD kink instability still comes into play because the poloidal field declines in the expanding flow faster than the toroidal field, but the instability develops in the already collimated flow so the jet preserves its directionality. Poynting energy flux should be converted into plasma energy. Gradual acceleration by magnetic forces (Vlahakis 2004; Vlahakis & Königl 2004a,b; Beskin & Nokhrina 2006; Komissarov et al. 2007) as well as dissipation of alternating magnetic fields (Levinson & van Putten 1997; Levinson 1998; Heinz & Begelman 2000) have also been proposed. While these processes could play a role in jets, magnetic energy release via the CD kink instability is inherent in narrow Poynting dominated jets. This instability disrupts the regular structure of the magnetic field to liberate magnetic energy (Lyubarskii 1992, 1999; Spruit et al. 1997; Begelman 1998;

Giannios & Spruit 2006), and naturally could result in flaring activity like that observed from the blazars.

3D RMHD simulations designed to investigate the properties of the CD kink instability of a static plasma column and more recently incorporating a sub-Alfvénic velocity shear surface (Mizuno et al. 2009; and work in progress) show that on the order of 100 Alfvén crossing times are required to reach the non-linear stages of kink development. In the SMBH jet collimation and acceleration environment an Alfvén crossing time is comparable to a light crossing time of the Poynting-flux spine. Of particular interest in the blazar context is the finding that a velocity shear surface inside some critical magnetic radius results in helically twisted flow within a growing and slowly moving kink as opposed to a kink embedded within and moving with the flow.

Jet expansion will significantly slow the spatial growth of a developing kink as the Alfvén crossing time increases. In this regard a recent fully 3D GRMHD simulation (McKinney & Blandford 2009) reveals that the jets generated by rotating, accreting black holes can be relatively stable structures out to $10^3$ gravitational radii and reach Lorentz factors of $\Gamma \sim 10$. Some mild twisting associated with a kink mode is evident in the simulation but no significant disruption or dissipation occurs. Beyond the Alfvén surface where the jet is super-Alfvénic there is a possibility of coupling between the CD kink and the Kelvin-Helmholtz (KH) velocity shear driven helical instability. Theory and simulations (Hardee 2007; Mizuno et al. 2007) have indicated that strong magnetic fields combined with a spine-sheath morphology can also slow the spatial growth of this mode. Thus, it should be possible for a jet to maintain a relatively strong and ordered helical magnetic field and flow structure through the blazar zone as needed to explain the Marscher et al. (2008, 2010) polarization results. Once the flow is super-magnetosonic, internally driven and externally driven shocks could form in the flow.

## 3. Microphysics inside the blazar zone

Inside the black hole magnetosphere and in the near jet generation zone the dominant particle acceleration processes should be associated with vacuum gap electric fields and magnetic reconnection. Electromagnetic models have been discussed by Levinson (2000), Maraschi & Tavecchio (2003), and Kundt & Gopal- Krishna (2004), and the consequences for emission by Krawczynski (2007). These particle acceleration processes cannot be studied on a fundamental level by MHD type numerical simulations but ultimately can be investigated on the microphysical level using general relativistic particle-in-cell (GRPIC) codes, e.g., Watson & Nishikawa (2009).

When the jet becomes super-Alfvénic and subsequently super-magnetosonic (location depending on magnetic reconnection at various scales, associated particle heating/acceleration and mass loading of the Poynting-flux spine) shocks can form. Typically we assume Fermi acceleration in the shock region, but shock models, e.g, Böttcher & Dermer (2010), have to assume an electron injection energy spectrum, a fraction, $\epsilon_B$, of the energy density in the shocked plasma contained in the magnetic field, and a fraction, $\epsilon_e$, of the shocked plasma kinetic energy transferred into relativistic electrons. These assumed microphysics parameters, whose values are usually fitted from the data, reflect our lack of understanding of the microphysics (Waxman 2006).

The microphysics, e.g., Spitkovsky (2008) and Nishikawa et al. (2009) from relativistic shocks is now being investigated using relativistic particle-in-cell (RPIC) codes,. Computational constraints have limited the majority of simulations to shocks in electron-positron pair plasmas and 2D simulations are used to cover parameter space combined with a few fully 3D simulations. In relativistic shocks the most rapid acceleration is provided in situ by the filamentation instability, so named as the interpenetrating plasmas form current filaments and associated magnetic fields. Particle acceleration occurs rapidly on electron plasma frequency timescales. For electron-proton plasmas the protons respond on proton plasma frequency timescales that are much longer than the electron plasma frequency timescales.

A 3D simulation of a Lorentz factor 15 unmagnetized pair jet penetrating an unmagnetized ambient pair plasma achieved partial development of a a hydrodynamic-like trailing shock, contact discontinuity, and leading shock structure. The strongest electromagnetic fields appeared in the nearly fully developed trailing shock with $\epsilon_B \sim 0.3$ (Nishikawa et al. 2009). 2.5D and 3D pair plasma shocks have been investigated in the contact discontinuity frame over a broad range of magnetic obliquities (Sironi & Spitkovsky 2009a). Fermi type particle acceleration was observed but with particle acceleration and spectral index sensitive to the magnetic obliquity and no "suprathermal" power law tail developed for magnetic fields not sufficiently oblique to the shock surface as significant self-generated shock turbulence could not develop upstream of the shock. However, very recent simulations investigating the effect of increasing the positive particle mass (private communication) indicate significant differences at sufficiently large "proton" masses, $m_p/m_e > 20$. Ultimately we may be able to tell the difference between pair plasma shocks and pair/proton plasma shocks and confirm or refute the one-zone blazar model requirement that jets are dynamically dominated by cold protons.

Other recent developments include the self-consistent computation of emission from pair plasma shocks (Sironi & Spitkovsky 2009b). In RPIC simulations, emission can be calculated self-consistently from the particle accelerations in the turbulent electric and magnetic fields accompanying the shock. These 2D based self-consistent pair plasma simulation results carried out in the reference frame of the shock contact discontinuity reveal synchrotron like self-absorbed spectra at lower frequencies, a thermal component, and higher frequency supra-thermal emission over a six order of magnitude frequency range.

Ultimately it is hoped that magnetized RMHD shock simulations on macroscopic scales can provide the global shock properties that can be inserted into coupled RPIC simulations to investigate the microphysics on much smaller scales. The results could provide a firm physical basis for magnetization and particle acceleration inputs into existing blazar shock models.

## 4. The case of M 87

### 4.1. The TeV-Radio Connection

The radio galaxy M 87 is a Fanaroff-Riley Type I source, the SED is similar to that of an LBL, and this suggests that M 87 is a misaligned BL Lac. The luminosity distance is D = 16.7 Mpc (Mei et al. 2007), so 1 mas corresponds to 0.081 pc = $2.5 \times 10^{17}$ cm in the sky plane. The mass of the central black hole is $(6.0 \pm 0.5) \times 10^9$ $M_\odot$ (Gebhardt & Thomas 2009), at the above distance. The Schwarzschild radius $R_s \sim 1.8 \times 10^{15}$ cm and 1 mas $\sim 140 R_s$. The most rapid optical proper motion is observed at HST-1 ($\sim$ 1 arcsec from the radio/optical core) with superluminal speed $\beta_{obs} \sim 6$ (Biretta, Sparks & Macchetto 1999). The most rapid radio proper motion at HST-1 indicates $\beta_{obs} \sim 4$ (Cheung, Harris, & Stawarz 2007). The limb brightened radio structure and proper motion difference has been interpreted as evidence for velocity shear.

The highest observed superluminal speed requires a jet viewing angle of $\theta < 19^o$ and the assumption of $\theta \sim 15^o$ seems reasonable. With this choice the intrinsic distance along the jet is $r_{int} \sim 3.86\, r_{obs}$, and 1 mas $\sim 540 R_s$ along the jet. The optical and radio proper motions then indicate Lorentz factors of $\sim 7.5$ and $\sim 4.5$, respectively at HST-1 at $\sim 5.3 \times 10^5 R_s$ along the jet.

VLBA observations at 43 GHz show a limb brightened jet to within 0.5 mas (270 $R_s$ along the jet) of the core. Wider opening angles are observed closer to the core (Junor et al. 1999; Walker et al. 2008; Kovalev et al. 2007). At 86 GHz the core is no larger than $25 \times 7\, R_s$ (Krichbaum, et al. 2006). The opening angle morphology and the observation of a faint counter-jet (Kovalev et al. 2007) are consistent with the jet converging at the central engine.

VHE gamma-ray emission was first reported by HEGRA in 1998/99 (Aharonian et al. 2003). The emission was confirmed by H.E.S.S. in 2003-2006 (Aharonian, et al. 2006a) with gamma-ray flux variability on time scales of days. During 50 nights between January and May 2008 H.E.S.S. (Aharonian et al. 2006b), MAGIC (Albert et al. 2008) and VERITAS (Acciari et al. 2008) found day scale variability in the energy range between 0.1 and 10's of TeV, and the flux reached the highest level ever observed from M 87, > 10% of the Crab Nebula.

Leptonic (Georganopoulos, Perlman & Kazanas 2005; Lenain 2008) and hadronic (Reimer, Protheroe & Donea 2004) VHE gamma-ray models have been proposed. The nucleus (Neronov & Aharonian 2007; Rieger & Aharonian 2008), the inner jet (Tavecchio & Ghisellini 2008) or HST-1, have been discussed as possible sites (Cheung, Harris & Stawarz 2007).

At X-ray frequencies during 2008 HST-1 is found in a low state, but the nucleus was found in its highest X-ray flux state since 2000 (Harris, Cheung & Stawarz 2009). This is in contrast to the 2005 VHE gamma-ray flares (Aharonian et al. 2006), which happened after a several year increase in the X-ray flux of HST-1 (Harris et al. 2006). Given its low X-ray flux in 2008, HST-1 is an unlikely site for the 2008 VHE flaring activity.

Throughout 2007 the VLBA observed M 87 at 43 GHz roughly every three weeks (Walker et al. 2008, 2009) and every 5 days from January to April 2008, see Acciari et al. (2009) with a resolution of $0.21 \times 0.43$ mas (resolution $\sim 230 R_s$ along the jet at a viewing angle of $15^o$). An initial radio flux density increase was located in the unresolved core. The radio peak occurred about 50 days after the onset of the VHE flare. Over this time the jet base brightened and extended about 0.77 mas, implying an average apparent velocity of 1.1 c, and Lorentz and Doppler factors of $\sim 1.8$ and $\sim 2.8$, respectively at $\theta = 15^o$.

### 4.2. Implications for the Blazar Zone

The observed emission can be explained by an ejection event in the central region associated with the VHE flare. Synchrotron self-absorption or the time needed to cool the electrons causes a delay of the observed radio peak. The radio structural change along with the timing of the VHE activity, imply that the VHE emission came from a region closer to the central engine than the VLBA resolution along the jet, $\sim 230 R_s$. The day scale variability implies an intrinsic timescale, $\Delta t_{int} << \delta_{max} \Delta t_{obs} \sim 15 \times 10^5$ sec, and acceleration and emission region size $<< 25 R_s = 4.5 \times 10^{16}$ cm $= c \Delta t_{int}$ where $\delta_{max} = 2\Gamma$ with $\Gamma = 7.5$.

TeV gamma-ray photons can escape the central region of M 87 without being heavily absorbed through $e^+ e^-$ pair production (Neronov & Aharonian 2007; Rieger & Aharonian 2008) with either photons from the accretion disk (Cheung, Harris & Stawarz 2007) or infrared photons as M 87 harbors a low luminosity accretion disk (Rieger & Aharonian 2008) and lacks a source of infrared radiation close to the central black hole (Perlman et al. 2007).

Acceleration by vacuum gap electric fields in the black hole magnetosphere (Neronov & Aharonian 2007) or due to centrifugal acceleration in an active plasma-rich environment, where the parallel electric field is screened (Rieger & Aharonian 2008) are definite possibilities. Synchrotron and curvature radiation of the charged particles, and inverse Compton scattering of thermal photons can produce VHE gamma-ray photons (Krawczynski 2007). An alternative scenario could be that the primary photons create a pair cascade whose leakage produces the observed gamma-ray emission (Bednarek 1997).

In an hadronic M 87 model (Reimer et al. 2004), a primary relativistic electron population is injected together with high-energy protons into a highly magnetized emission region. The VHE emission is dominated by either $\mu^\pm / \pi^\pm$ synchrotron radiation or by proton synchrotron radiation. The low-energy component is explained by the synchrotron emission of the electron population. The requirement of magnetic fields $\sim 10^3$ Gauss is inconsistent with the low accretion flow, low density and low magnetic field environment inferred for M 87 (see Neronov & Aharonian 2007), although not ruled out during short episodes. This makes an hadronic origin for the TeV emission unlikely.

Neronov & Aharonian (2007) demonstrate that the TeV emission from M 87 can be explained as IC emission of ultrarelativistic $e^+e^-$ pairs produced in an electromagnetic cascade in the black hole magnetosphere with acceleration and radiation occurring in the magnetospheric vacuum gap. With acceleration and emission processes occurring on spatial scales on the order of $R_s$, the day scale TeV variability that requires $c\Delta t \sim R_s$ for a stationary acceleration and emission region is satisfied. Thus, a magnetospheric vacuum gap model for the TeV emission remains a definite possibility for M 87.

One-zone blazar models would appear to require unrealistic Doppler factors in the M 87 jet beyond the radio core (Tavecchio & Ghisellini 2007). This difficulty is remedied by the adoption of a spine-sheath structure such as is suggested by proper motions at HST-1, implied by the limb brightened subparsec scale jet, and on the basis of theoretical arguments (e.g., Henri & Pelletier 1991) and GRMHD numerical simulations (e.g., Hawley & Krolik 2006; McKinney 2006). Tavecchio & Ghisellini (2007) are able to produce reasonable fits to the M 87 SED using spine-sheath Lorentz factors and a viewing angle not significantly different from those appropriate to optical and radio proper motions at HST-1.

The rapid time variability of the TeV emission poses a difficulty if a spine-sheath model is applied to the jet as a whole. A similar problem arises if particle acceleration is the result of magnetic reconnection applied to the jet as a whole. With a reconnection timescale on the order of 100 Alfvén-light crossing times (Mizuno et al. 2009), particle acceleration can only be associated with reconnection occurring on spatial scales $<< 25R_s/100$.

Thus, smaller scale "needles" within the jet seem needed to account for rapid time variability and particle acceleration by magnetic reconnection. In this regard Lenain et al. (2008) showed that a high-energy emission region (X-rays up to VHE) could consist of small blobs ($\sim 10^{14}$ cm) moving within the large opening angle jet base [similar in spirit to the needles within a jet model] and radiating at distances just beyond the Alfvén surface, $\sim 100R_s$. The fast, compact blobs could contribute to X-rays and gamma-rays through the synchrotron self-Compton mechanism, and be embedded in an extended, diluted and slower jet emitting synchrotron radiation from radio to optical frequencies. A slow rise of radio emission is expected because the emission volume is synchrotron self-absorbed at radio frequencies. Here it should be noted that TeV emission coming from within $\sim 230R_s$ along the jet can come from the broad jet base, full intrinsic opening angel of $\sim 17^o$ at $\theta = 15^o$, where part of flow could lie at less than $7^o$ to the line of sight with the jet axis at $15^o$ to the line of sight.

The "jet-in-a-jet" model of Giannios et al. (2009) remains a possibility for the fast TeV variability as in this model high Lorentz factor structures within the Poynting flux region could lie within $230R_s$ along the jet and could again come from the broad jet base. A rapid jet deceleration model also remains a possibility as the proposed jet deceleration zone lies within $230R_s$ along the jet (Georganopoulos et al. 2005) and the background radiation field is of the correct order to give rise to the required deceleration but still be transparent enough to TeV gamma-rays (Levinson 2007). In these models the large scale flow Lorentz factor beyond $230R_s$ along the jet need not significantly exceed the Lorentz factor at HST-1.

The mm radio core in M 87 seems to be nearly coincident with the central engine and well inside the radio core distance from the central engine in PKS 1510-089 (Marscher et al. 2010) or even BL Lac (Marcher et al. 2008). This may be a result of the relatively large viewing angle and reduced synchrotron self absorption. However, we note that the jet is highly structured out to large distances from the core and there is evidence for helically twisted filaments along the kiloparsec jet (Lobanov et al. 2003). This provides circumstantial evidence for helically ordered flow and magnetic fields on the parsec scales suggested by the multiwavelength studies of PKS 1510-089 and BL Lac. On the other hand, there is no evidence for a recollimation shock in the M 87 jet downstream from the radio core. This fact may simply mean that recollimation shocks are not always present in AGN jets or coincident with the radio core.

## 5. Conclusions

The results from GRMHD jet generation simulations along with the limb brightened radio structure of M 87 at sub-parsec and kiloparsec scales suggests that a high speed spine and lower speed sheath morphology is likely to exist within the blazar zone. These results suggest that we can decouple the synchrotron and IC regions and reduce the Doppler factors required by simple homogeneous one-zone blazar models. GRMHD simulation results along with theoretical and numerical CD and KH jet instability work indicate that a jet can be sufficiently stable but at the same time contain helically organized flow and magnetic fields within the blazar zone.

For M 87, observed proper motions at sub-parsec to a few hundred parsec (HST-1) scales indicate superluminal sheath motions from 1 - 2 c (sub-parsec) to 4 c (HST-1) and superluminal spine motions of 6 c (HST-1). At a viewing angle $\theta = 15^o$ these results suggest sheath Lorentz factors from 2 (sub-parsec) to 4.5 (HST-1) and, at least at HST-1, a BL Lac like spine Lorentz factor of 7.5. At $\theta = 15^o$ these Lorentz factors imply only modest deboosting of the spine relative to the sheath. Higher spine Lorentz factors and/or an intrinsically radio faint spine are required to achieve the observed radio limb brighten-ing. Higher spine Lorentz factors cannot be ruled out on the sub-parsec jet.

The VLBI jet collimation morphology and faint counter-jet suggest that the radio core is nearly coincident with the central black hole. Here the broad jet base allows a part of the jet flow to be pointed within $7^o$ of the line of sight.

Outside the M 87 radio core the jet expands uniformly all the way to knot A (a few kiloparsecs). This indicates that conversion of Poynting flux to kinetic flux occurs inside $230 R_s$ along the jet. CD instability driven jet spine magnetic reconnection and reconfiguration remains a possibility on these spatial scales. This is consistent with an Alfvén point on the order of $100 R_s$ from the central black hole, and the uniformity of the radio structure from sub-parsec to kiloparsec scales indicates that a recollimation shock need not exist within the blazar zone.

The observed SED for M 87 seems adequately described by leptonic models and hadronic models for emission coming from inside $230 R_s$ along the jet are unlikely. This result can likely be extended to BL Lacs with low accretion rates and correspondingly weak magnetic fields.

Like other TeV blazars the TeV variability requires relatively small acceleration and emission regions. In the case of M 87 this suggests vacuum gap electric field particle acceleration in the black hole magnetosphere, local magnetic reconnection driven particle acceleration on the small scale size of the "jet-in-a-jet" model, or rapid shock acceleration in the "needle" type model.

At least for M 87 there appears to be a definite connection between the TeV, possibly X-ray, and radio events associated with a central engine ejection episode.

*Acknowledgements.* P. Hardee acknowledges research support via NASA-ATFP award NNX08AG83G and NSF award AST-0908010.